\documentclass[11pt,twoside]{article}
\usepackage{asp2014}

\aspSuppressVolSlug
\resetcounters

\begin{document}

\title{Spatial Variations of the Interstellar Polarization and Interstellar Extinction}
\author{George~Gontcharov,$^1$
\affil{$^1$Central (Pulkovo) Astronomical Observatory, Russian Academy of Sciences, 
65/1 Pulkovskoye Shosse, Saint-Petersburg, 196140 Russia; \email{georgegontcharov@yahoo.com}}}

\paperauthor{George~Gontcharov}{georgegontcharov@yahoo.com}{orcid.org/0000-0002-6354-3884}{Central (Pulkovo) Astronomical Observatory}{Laboratory of the Dynamics of the Galaxy}{Saint-Petersburg}{}{196140}{Russia}

\begin{abstract}
For more than 5000 stars with accurate parallaxes from the Hipparcos and
Gaia DR1 Tycho-Gaia astrometric solution (TGAS),
Tycho-2 photometry, interstellar polarization from eight catalogues and interstellar extinction
from eight 3D maps the largest up to date comparison of the polarization and extinction is provided.
The polarization catalogues are:
Heiles (2000),
Berdyugin et al. (2014),
Santos et al. (2011),
Berdyugin et al. (2001),
Berdyugin \& Teerikorpi (2002),
Leroy (1999),
Cotton et al. (2016),
Wills et al. (2011).
The direct comparison of the data from these catalogues for common stars shows that the data are free
from considerable systematic errors and can be used together.
The eight maps of the extinction are described by Gontcharov \& Mosenkov (2016).
The maps give different estimations of the extinction, the polarization efficiency as the polarization divided
into extinction $P/A_V$ as well as the percentage of the stars with the polarization efficiency higher than
the limit of Serkowski et al. (1975) $P/A_V>0.03$.
Using the Hipparcos parallaxes we see about 200 stars (4\%, mainly OB stars)
drop higher than the limit when we use any extinction map.
This percentage and list of the stars is similar to those by other authors.
However, using the TGAS parallaxes we see only 17 such stars (0.3\%).
They shoud be considered as the stars with probable own (non-interstellar) polarization.
The stars overcoming the limit with the Hipparcos parallaxes but not with TGAS ones have the latter
smaller (distances are larger) than the former.
Therefore, those stars are behind some distant dense dust clouds and have higher extinction, higher polarization
but usual polarization efficiency.
The spatial variations of the extinction and polarization are in agreement.
The most explicit is the dependence of the extinction, polarization and its efficiency on distance
shown in Figure 1.
The polarization and extinction are negligible inside the Local Bubble within 80 pc from the Sun.
In the vast Bubble's shell at the distances 80--118 pc from the Sun the polarization and extinction
rapidly grow with the distance whereas the position angle of the polarization is oriented predominantly
along the shell of the Bubble. Outside the Bubble the polarization and extinction grow with the
distance slowly.
In addition, within a radius of 80--300 pc of the Sun a disc of some filamentary dust clouds
(including well-known Markkanen (1979) cloud) is observed as in the polarization map as in the reddening
one by Schlegel et al. (1998).
This disk is inclined to the galactic plane at an angle of about 45 deg and passing through the Sun.
In this disc the position angle of polarization is preferably oriented along the plane of the disk.
For the regions further than 300 pc the position angle of polarization is preferably oriented along the
Local spiral arm, i.e. Y coordinate axis.
The polarization and its efficiency is lower in the dust layer in the Gould belt than in the equatorial
dust layer. It may means different properties of dust in these two layers.
\end{abstract}

\acknowledgements The study was financially supported by the
``Transient and Explosive Processes in Astrophysics''
Program P-7 of the Presidium of the Russian Academy of Sciences.

\articlefigure{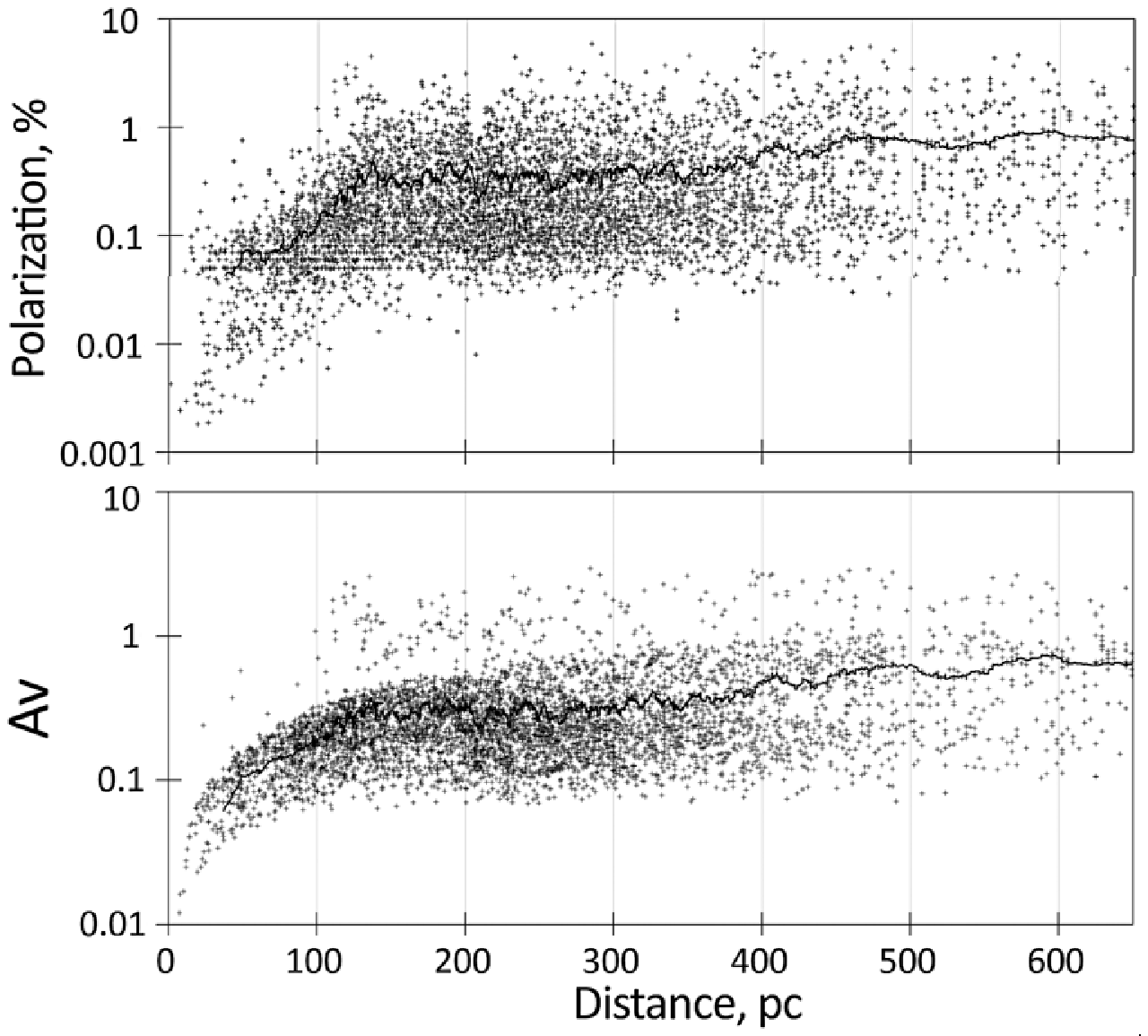}{g31}{The polarization and extinction 
in dependence on the distance. 
The curves are the moving averaging over 77 points.}

\end{document}